\documentstyle[aps,pra,amsmath,twocolumn,epsfig]{revtex}

\begin{document}
\draft
\title{Landau damping of transverse quadrupole oscillations of an elongated
Bose-Einstein condensate}
\author{M. Guilleumas$^1$ and L. P. Pitaevskii$^{2,3}$}
\address{$^1$Dept. d'Estructura i Constituents de la Mat\`eria,\\
Facultat de F\'{\i}sica,\\
Universitat de Barcelona, E-08028 Barcelona, Spain}
\address{$^2$Dipartimento di Fisica, Universit\`a di Trento and\\
Istituto Nazionale per la Fisica della Materia, I-38050 Povo, Italy}
\address{$^3$Kapitza Institute for Physical Problems, 117454 Moscow,\\
Russian Federation}
\date{\today}
\maketitle

\begin{abstract}
We study the interaction between low-lying transverse collective
oscillations and thermal excitations of an elongated Bose-Einstein
condensate by means of perturbation theory. We consider a cylindrically
trapped condensate and calculate the transverse elementary excitations at
zero temperature by solving the linearized Gross-Pitaevskii equations in two
dimensions. We use them to calculate the matrix elements between thermal
excited states coupled with the quasi-2D collective modes. The Landau
damping of transverse collective modes is investigated as a function of
temperature. At low temperatures, the damping rate due to the Landau decay
mechanism is in agreement with the experimental data for the decay of the
transverse quadrupole mode, but it is too small to explain the slow
experimental decay of the transverse breathing mode. The reason for this
discrepancy is discussed.
\end{abstract}

\pacs{03.75.Fi, 02.70.Lq, 67.40.Db}

\section{Introduction}

Transverse collective oscillations have been recently excited in
the radial plane of an elongated cylindrically symmetric
condensate \cite{Chevy01}. They are quasi-2D excitations and
indeed, at very low temperatures, the measured frequencies of the
transverse monopole and quadrupole modes are found to be in very
good agreement with the theoretical predictions obtained by
linearizing the Gross-Pitaevskii equation for a trapped
two-dimensional (2D) system. Moreover, it has been experimentally
found that the transverse breathing mode of an elongated
condensate exhibits unique features. First, the excitation
frequency is close to twice the radial trapping frequency, it is
almost independent of the strength of the two-body interaction and
the number of particles, and it is nearly independent of
temperature. And second, the transverse breathing mode has a very
small damping rate compared to other modes. The measured damping
rate of the transverse quadrupole mode is approximately one order
of magnitude larger than the one of the transverse breathing mode.
This is in contrast to the 3D case, where both monopole and
quadrupole modes have similar decay rates. These striking
properties experimentally found for the transverse breathing mode
are in agreement with the unique features predicted for the
breathing mode in 2D isotropic condensates
\cite{Pit96,Kagan96,Pitaevskii97}: its frequency is a universal
quantity and it is an undamped mode. One can conclude that these
features are characteristic of the reduced dimensionality thus,
providing a demonstration of the two-dimensional nature of the
transverse oscillations of an elongated condensate. Therefore, a
highly anisotropic cigar-shaped trap can be considered as a first
approximation as an infinite cylinder along the $z$-direction.

In the present paper we consider a cylindrical condensate and
calculate at zero temperature the transverse spectrum of
excitations, by solving the linearized Gross-Pitaevskii equation
in 2D, and the modes with non-zero momentum $k$ along the
longitudinal axis, as well. The main purpose is to investigate the
coupling between quasi-2D (transverse) collective oscillations and
thermal excitations in the collisionless regime. We study the
Landau damping as a decay mechanism of transverse collective modes
as a function of temperature, in which a transverse collective
oscillation is annihilated in a collision with a thermal
excitation to give rise to another excitation.

This paper is organized as follows. In Sec.~II we calculate the
transverse spectrum of excitations of a cylindrical condensate by
solving the linearized Gross-Pitaevskii equation in 2D within the
Bogoliubov theory. We obtain also the modes with non-zero $k$. In
Sec.~III we briefly recall the perturbation theory for the
interaction between collective modes of a condensate and thermal
excitations (developed in Ref.~\cite{Pitaevskii97PLA}). We apply
this theory to calculate the decay rate due to Landau damping of a
transverse oscillation. In Sec.~IV we discuss the main results for
the transverse quadrupole mode. The situation with the singular
transverse breathing mode is also discussed.

\section{Elementary excitations of a cylindrically trapped condensate}

At low temperature, the elementary excitations of a trapped weakly
interacting degenerate Bose gas are described by the
time-dependent Gross-Pitaevskii (GP) equation for the order
parameter:
\begin{equation}
i\hbar {\frac{\partial}{\partial t}} \Psi ({\bf r},t) = \left( - {\ \frac{%
\hbar^2 \nabla^2}{2M}} + V_{{\rm ext}}({\bf r}) + g \mid \!\Psi({\bf r}%
,t)\!\mid^2 \right) \Psi({\bf r},t) \; ,  \label{TDGP}
\end{equation}
where $g=4\pi \hbar^2 a/M$ is the interaction coupling constant,
fixed by the $s$-wave scattering length $a$, $M$ is the atomic
mass and the order parameter is normalized to the number of atoms
in the condensate $N$. The confining potential is usually
cylindrically symmetric and it is given by
$V_{{\rm ext}}(r_\perp,z)=M(\omega_{%
\perp}^2 r_\perp^2+\omega_z^2 z^2)/2$, with $r_\perp^2=x^2+y^2$.
The ratio between axial ($\omega_z$) and radial ($\omega_{\perp}$)
trapping frequencies defines the anisotropy parameter of the trap
$\lambda = \omega_z/\omega_\perp$.

Experimentally, in Ref.~\cite{Chevy01}, the transverse modes have
been
excited in the radial plane of a highly anisotropic cigar-shaped trap, with $%
\lambda \sim 0.0646$. Thus, as a reasonable approximation, we
consider an idealized cylindrical trap that is uniform in the
$z$-direction ($\omega _{z}=0$) and has isotropic trapping
potential in the radial plane
\begin{equation}
V_{ext}(r_{\perp })=\frac{1}{2}M\omega _{\perp }^{2}r_{\perp }^{2}\,.
\label{Vtrap}
\end{equation}
The harmonic trap frequency $\omega _{\perp }$ provides a typical
length scale for the system, $a_{\perp }=(\hbar /M\omega _{\perp
})^{1/2}$. We assume a cylindrical condensate with large
longitudinal size $L$ compared to the radial size $R$, and with a
number of atoms per unit length $N_{\perp}=N/L$.

At low temperatures, the dynamics of the condensate is described by the
linearized time-dependent GP equation. Assuming small
oscillations of the order parameter around the ground-state value
\begin{equation}
\Psi({\bf r},t)=e^{-i\mu t/\hbar} \left[ \Psi_0(r_\perp) + \delta \Psi({\bf r%
},t) \right] \,,  \label{fluctuation}
\end{equation}
where $\mu$ is the chemical potential,
the normal modes of the condensate can be found by seeking fluctuations such
as
\begin{equation}
\delta \Psi({\bf r},t)=u({\bf r}) e^{-i \omega t} +v^*({\bf r}) e^{i \omega
t} \,.  \label{deltaPsi}
\end{equation}
The functions $u$ and $v$ are the ``particle" and ``hole" components
characterizing the Bogoliubov transformations, and $\omega$ is the frequency
of the corresponding excited state.

The transverse excitations of a highly elongated condensate can be obtained
by neglecting oscillations along the longitudinal direction and thus,
looking for fluctuations in the transverse plane of the form \cite
{Stringari98,Ho99}
\begin{equation}
\delta \Psi(r_\perp,\theta,t)=U(r_\perp) \, e^{i m \theta} e^{-i \omega
t}+V^*(r_\perp) \, e^{-i m \theta} e^{i \omega t} \,,  \label{deltaPsitrans}
\end{equation}
where we have used cylindrical coordinates ($r_\perp,\theta,z$).

Transverse excitations are quasi-2D modes. To calculate Landau damping we
must also consider thermal excitations having a quasi-continuous spectrum at
$L\gg R$ with the finite longitudinal wave vector $k$ along $z$ direction.
To study these modes we look for solutions that correspond to propagating
waves of the form \cite{Stringari98,Zaremba98}
\begin{eqnarray}
\delta \Psi ({\bf r},t) &=&U_{k}(r_{\perp })\,e^{i(m\theta +kz)}e^{-i\omega
t}+  \nonumber \\
&&V_{k}^{\ast }(r_{\perp })\,e^{-i(m\theta +kz)}e^{i\omega t}\,.
\label{deltaPsilong}
\end{eqnarray}

Inserting (\ref{fluctuation}), (\ref{deltaPsi})
and (\ref{deltaPsilong}) in Eq.~(\ref{TDGP})
and retaining terms up to first order in $u$ and $v$, it follows three
equations. The first one is the nonlinear equation for the order parameter
of the ground state,
\begin{equation}
\left[ H_{0}+g\Psi _{0}^{2}(r_{\perp })\right] \Psi _{0}(r_{\perp })=\mu
\Psi _{0}(r_{\perp })\,,  \label{groundstate}
\end{equation}
where $H_{0}=-(\hbar ^{2}/2M)\nabla _{r_{\perp }}^{2}+V_{{\rm ext}}(r_{\perp
})$, and $\nabla _{r_{\perp }}^{2}=d^{2}/dr_{\perp }^{2}+r_{\perp
}^{-1}d/dr_{\perp }$; while $U_{k}(r_{\perp })$ and $V_{k}(r_{\perp })$ obey
the following coupled equations~\cite{Dalfovo99}:
\begin{eqnarray}
\hbar \omega U_{k}(r_{\perp }) &=&\left[ H_{0}+\frac{\hbar ^{2}}{2M}\left(
\frac{m^{2}}{r_{\perp }^{2}}+k^{2}\right) -\mu +2g\Psi _{0}^{2}\right]
U_{k}(r_{\perp })  \nonumber \\
&&+g\Psi _{0}^{2}V_{k}(r_{\perp })  \label{coupled1} \\
-\hbar \omega V_{k}(r_{\perp }) &=&\left[ H_{0}+\frac{\hbar ^{2}}{2M}\left(
\frac{m^{2}}{r_{\perp }^{2}}+k^{2}\right) -\mu +2g\Psi _{0}^{2}\right]
V_{k}(r_{\perp })  \nonumber \\
&&+g\Psi _{0}^{2}U_{k}(r_{\perp })\;.  \label{coupled2}
\end{eqnarray}
We solve the linearized GP equations Eqs.~(\ref{groundstate}-%
\ref{coupled2}) at zero temperature, to obtain the ground state wave
function $\Psi _{0}$, the eigenfrequencies $\omega (k)$ and hence the
energies $E=\hbar \omega $ of the excitations, as well as the corresponding
functions ($u,v$).

In a cylindrical trap the modes of the condensate are labeled by $i=(n,m,k)$%
, where $n$ is the number of nodes in the radial solution, $m$ is the $z$%
-projection of the orbital angular momentum, and $k$ is the longitudinal
wave vector. The quantum numbers $n$ and $m$ are discrete ($n=0,1,2,...$, and
$m=0,\pm 1,\pm 2, ....$) whereas $k$ for highly elongated condensates ($L\gg
R$) is a real number, and thus, it is a continuous index.
Moreover, it can be seen
from Eqs.~(\ref{coupled1}) and (\ref{coupled2}), that levels with $m=\pm\mid
m \mid$ or $k=\pm \mid k \mid$ are exactly degenerated.

The excitation spectrum of a cylindrical condensate presents different
branches labeled by ($m,n$) that are continuous and increasing functions of $%
k$ \cite{Zaremba98}. For a given mode, i.e., fixed $n$ and $m$, the
frequency of excitation is $\omega (k)$.

We have numerically checked from Eqs.~(\ref{coupled1}) and (\ref{coupled2})
that the state with ($m=0,n=1$) and $k=0$ corresponds to the transverse
breathing mode with excitation frequency equal to $2\omega _{\perp }$, and
that it is independent of the strength of the two-body interaction potential
and the number of particles. Of course, according to the universal nature of
this mode noted above,
this general result is also valid in
the Thomas-Fermi (TF) limit, where the dispersion relation can be found
analytically\cite{Stringari98,Ho99}.
However, the functions $u$ and $v$ for this mode do not coincide in
general with their Thomas-Fermi limit.

The state with ($m=2,n=0$) and $k=0$ is a transverse quadrupole mode whose
frequency depends on the two-body interaction, ranging from the TF value $%
\sqrt{2}\omega _{\perp }$, valid for large condensates
($aN_{\perp}\gg 1$ or equivalently $\mu \gg \hbar
\omega_{\perp}$), to the noninteracting value $2\omega _{\perp }$.
Therefore, the transverse quadrupole has to be obtained
numerically by solving Eqs.~(\ref{coupled1}) and (\ref{coupled2})
in each particular value of the dimensionless parameter $aN_{\perp
}$. In the present work, we use the numerical solutions to
calculate the Landau damping of collective modes.

It is well known that the dipole mode ($m=1,n=0$) is unaffected by
two-body interactions due to the translational invariance of the
interatomic force which cannot affect the motion of the center of
mass \cite{Dalfovo99} and its excitation frequency is $1
\omega_\perp$. It is worth stressing that the independence of the
interaction of the transverse monopole mode, is a unique property
of 2D systems, related to the presence of a hidden symmetry of the
problem described by the two-dimensional Lorentz group SO(2,1)
\cite{Pitaevskii97}.

\section{Landau damping}

Let us consider a collective mode with frequency $\Omega _{{\rm osc}}$, and
the excited states $i,j$ available by thermal activation, with energies $%
E_{i}$ and $E_{j}$, respectively. Suppose that this collective mode has been
excited and, therefore, the condensate oscillates with the corresponding
frequency $\Omega _{{\rm osc}}$. Due to interaction effects, the thermal
cloud of excitations can either absorb or emit quanta of this mode producing
a damping of the collective oscillation. We want to study the decay process
in which a quantum of oscillation $\hbar \Omega _{{\rm osc}}$ is annihilated
(created) and the $i$-th excitation is transformed into the $j$-th one (or
viceversa). The energy is conserved during the transition process, therefore
\begin{equation}
E_{j}=E_{i}+\hbar \Omega _{{\rm osc}}\,,  \label{selection}
\end{equation}
where we assume that $E_{j}>E_{i}$. This mechanism is known as Landau
damping. Another possible decay mechanism, also due to the coupling between
collective and thermal excitations, is the Beliaev damping \cite{Beliaev},
which is based on the decay of an elementary excitation into a pair of
excitations.

Let us define the dissipation rate $\gamma $ through the following relation
between the energy of the system $E$ and its dissipation $\dot{E}$:
\begin{equation}
\dot{E}=-2\gamma E\,.  \label{dampdef}
\end{equation}
Assuming that the damping is small, one can use perturbation theory to
calculate
the probabilities for the transition between the $i$-th excitation and the $%
j $-th one, both available by thermal activation, yielding the following
expression for the Landau damping rate \cite{Pitaevskii97PLA,Guilleumas00}:
\begin{equation}
\frac{\gamma }{\Omega _{{\rm osc}}}=\sum_{ij}\gamma _{ij}\,\delta (\omega
_{ij}-\Omega _{{\rm osc}})\,,  \label{damping2}
\end{equation}
where the transition frequencies $\omega _{ij}=(E_{j}-E_{i})/\hbar $ are
positive, and the delta function ensures the energy conservation during the
transition process. We have assumed that the thermal cloud is at
thermodynamic equilibrium and the states $i,j$ are thermally occupied with
the usual Bose factor $f_{i}=[\exp (E_{i}/k_{B}T)-1]^{-1}$. The ``damping
strength'' has the dimensions of a frequency and is given by
\begin{equation}
\gamma _{ij}=\frac{\pi }{\hbar ^{2}\Omega _{{\rm osc}}}\mid A_{ij}\mid
^{2}(f_{i}-f_{j})\,.  \label{dampingik}
\end{equation}
The matrix element that couples the low-energy collective mode ($u_{{\rm osc}%
},v_{{\rm osc}}$) with the higher energy single-particle excitations (for
which we use the indices $i,j$) is \cite{Pitaevskii97PLA}
\begin{eqnarray}
A_{ij} &=&2g\int d{\bf r}\,\Psi _{0}[(u_{j}^{\ast }v_{i}+v_{j}^{\ast
}v_{i}+u_{j}^{\ast }u_{i})u_{{\rm osc}}  \nonumber \\
&&+(v_{j}^{\ast }u_{i}+v_{j}^{\ast }v_{i}+u_{j}^{\ast }u_{i})v_{{\rm osc}}].
\label{matrixel}
\end{eqnarray}

In this work we calculate the quantities $\gamma_{ij}$ by using
the numerical solutions $u$ and $v$ of
Eqs.~(\ref{groundstate}-\ref{coupled2}) into the integrals
(\ref{matrixel}), avoiding the use of further approximations in
the spectrum of excitations.
In cylindrically symmetric traps, $%
i=(n,m,k)$ and from (\ref{deltaPsilong}) the longitudinal excitations are $%
u_{n\,m\,k}({\bf r})=U_{n\,m\,k}(r_\perp) \,e^{i m \theta} e^{i k
z}$ and the energies $E_{n\,m\,k}$ are not degenerate.

We are interested in the decay of transverse collective
excitations (\ref {deltaPsitrans}) $u_{{\rm osc}}(r_{\perp
},\theta )=U_{{\rm osc}}(r_{\perp })e^{im_{{\rm osc}}\theta }$,
due to the coupling with thermally excited states ($i,j$). The
functions $u_{{\rm osc}}$ and $v_{{\rm osc}}$ do not depend on
$z$, and from Eq.~(\ref{matrixel}) it is straightforward to see
that the matrix element $A_{ij}$ couples only energy levels
($i,j$) with the same quantum number $k\equiv k_{i}=k_{j}$, and
$\Delta m\equiv \mid m_{i}-m_{j} \mid =m_{{\rm osc}}$. To
calculate the damping rate $\gamma$ of the collective oscillation
we have to sum up over all couples of excited states ($i,j$), or
equivalently, over all pairs of sets
($n_{i}\,m_{i}\,k\,,n_{j}\,m_{j}\,k$),
that satisfy the above conditions. 
Since $k$ is a continuous quantum number, the sum over $k$ becomes and
integral and Eq.~(\ref{damping2}) yields
\begin{equation}
\frac{\gamma }{\Omega _{{\rm osc}}}=\sum_{n_{i}n_{j}m_{i}m_{j}}\int_{-\infty
}^{\infty }\frac{dk}{2\pi }\,\gamma _{ij}\,\delta (\omega _{ij}-\Omega _{%
{\rm osc}})\,,  \label{gamma2}
\end{equation}
where the sum is restricted to pairs that satisfy $\Delta m=m_{{\rm osc}}$.
The frequency of a mode is a continuous function of $k$, therefore the
transition frequency is $\omega _{ij}=\omega _{ij}(k)$. Then, in Eq.~(\ref
{gamma2}) the integration over $k$ yields
\begin{equation}
\frac{\gamma }{\Omega _{{\rm osc}}}=\frac{1}{2\pi }\sum_{\tilde{k}%
}\sum_{n_{i}n_{j}m_{i}m_{j}}\gamma _{ij}\,\left[ \,\left| \frac{\partial
\omega _{i}}{\partial k}-\frac{\partial \omega _{j}}{\partial k}\right| _{%
\tilde{k}}\right] ^{-1}  \label{gamma3}
\end{equation}
where $\tilde{k}$ is the wave vector in which the conservation of
energy is verified, i.e., $\mid \omega _{i}(\tilde{k})-\omega
_{j}(\tilde{k})\mid =\Omega _{{\rm osc}}$, and we have used the
known property of delta functions
\begin{equation}
\int_{-\infty }^{\infty }dk\,\delta (g(k))=\sum_{\tilde{k}}\frac{\delta (k-%
\tilde{k})}{\mid g^{\prime }(\tilde{k})\mid }\,.
\end{equation}

\section{Results}

We want to calculate the characteristic decay rates due to Landau damping of
the transverse low-lying collective modes of an elongated Bose-Einstein
condensate as a function of $T$. We consider the collective excitations in
the collisionless regime, which is achieved at low enough temperatures.

For a fixed number of trapped atoms, the number of atoms in the condensate
depends on temperature. At zero temperature the quantum depletion is
negligible and all the atoms can be assumed to be in the condensate \cite
{Dalfovo97}. At finite temperature the condensate atoms coexist with the
thermal bath. However, at low enough temperatures the excitation spectrum
can be safely calculated by neglecting the coupling between the condensate
and thermal atoms \cite{Dalfovo99}. It means that the excitation energies at
a given $T$ can be obtained within Bogoliubov theory at $T=0$ normalizing
the number of condensate atoms to the corresponding condensate fraction at
that temperature and assuming that they are thermally occupied with the Bose
factor.

In order to present numerical results we choose a gas of $^{87}$Rb atoms
(scattering length $a=5.82\times 10^{-7}$ cm) confined in an elongated
cylindrical trap with frequencies $\omega _{z}\approx 0$ and $\omega _{\perp
}=219\times 2\pi $ Hz, that corresponds to an oscillator length $a_{\perp
}=0.729\times 10^{-4}$~cm. The number of condensate atoms per unit length is
taken to be $N_{\perp }=N/L=2800\,a_{\perp }^{-1}$. These conditions are
close to the experimental ones of Ref.~\cite{Chevy01}.

We have solved the linearized GP equations to obtain an exact
description of the ground state $\Psi _{0}(r_{\perp })$ and the
normal modes of the condensate within Bogoliubov theory without
using the Thomas-Fermi or Hartree-Fock approximations. We have
obtained the following numerical results: $\mu =9.526\,\hbar
\omega _{\perp }$, and the excitation frequencies of the
transverse monopole $\Omega _{M}=2.0\,\omega _{\perp }$, and of
the transverse quadrupole mode $\Omega _{Q}=1.436\,\hbar \omega
_{\perp }$. We have calculated the branches of the excitation
spectrum of the cylindrical condensate, labeled by ($m,n$) as a
function of $k$.

To calculate the damping rate of a collective mode $\Omega_{{\rm
osc}}$ we have to obtain, first, all pairs of levels ($i,j$) of
the excitation spectrum that satisfy both the energy conservation
of the transition process $E_j-E_i=\hbar \Omega_{{\rm osc}}$, and
the selection rules given by the particular collective mode under
study. And then, calculate the corresponding matrix elements. We
have restricted our calculations to levels with energy $E \leq 7
\mu$. The contribution of higher excited levels can be neglected
since their occupation becomes negligible in the range of
temperatures we have considered.

It is worth recalling that the dipole mode is undamped \cite{Dalfovo99}
since it is not affected by two-body interactions. We have checked that
within our formalism the Landau damping of the transverse dipole mode is
zero, as expected.

First of all, we have calculated the Landau damping of the
transverse quadrupole mode ($k=0$, $m=2$ and $n=0$) as a function
of $T$, $\gamma _{Q}(T)$. For such a mode, the functions $u_{{\rm
osc}}$ and $v_{{\rm osc}}$ depend on the radial coordinate
$r_{\perp }$ and have also angular dependence ($\sim e^{\pm
i2\theta }$). Therefore, from Eq.~(\ref{matrixel}) it follows the
selection rules corresponding to transverse quadrupole-like
transitions: $\Delta m=\mid m_{j}-m_{i}\mid =2$ and $\Delta k=0$.
Only the energy levels ($i,j$) that satisfy these conditions can
be coupled by the interaction. The integration of the radial part
has to be done numerically.

Fixed the number of condensate atoms, we have calculated the
Landau damping for the transverse quadrupole mode as a function of
temperature. At zero temperature we have obtained the excitation
spectrum within Bogoliubov theory. We have found that there are
$34$ pairs of levels that verify the transverse quadrupole
selection rules and the energy condition (\ref{selection}) at
finite wave vectors, i.e., at $\tilde{k}\neq 0 $. We have
calculated the matrix element $A_{ij}$ for each allowed
transition, and then, the corresponding damping strength $\gamma
_{ij}$ at a given $T$. Summing up the damping strengths for all
transitions, we have obtained the damping rate for the transverse
quadrupole oscillation, due to the coupling with thermal
excitations.

\begin{figure}
\centerline{\epsfig{figure=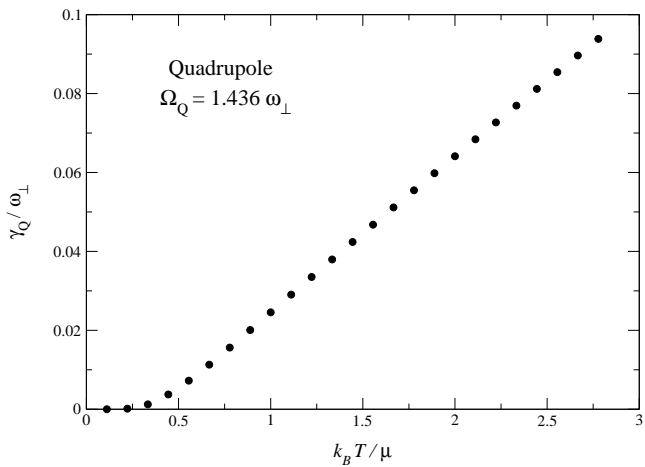,width=7.5cm,angle=-90}}
\caption{
Dimensionless damping rate of the transverse quadrupole mode $%
\protect\gamma_{{\rm Q}}/\protect\omega_\perp$ as a function of $k_B T/%
\protect\mu$ for a cylindrical condensate with $N_\perp=2800 a_\perp^{-1}$%
atoms of $^{87}$Rb per unit length. }
\label{fig1}
\end{figure}

In Figure \ref{fig1} we plot the damping rate versus $k_{B}T/\mu $
for the transverse quadrupole mode for a fixed number of
condensate atoms per unit length in a cylindrical trap. As
expected, Landau damping increases with temperature since the
number of excitations available at thermal equilibrium is larger
when $T$ increases. One can distinguish two different regimes, one
at very low $T$ ($k_{B}T\ll \mu $) and the other at higher $T,$
where damping is linear with temperature. However, this linear
behaviour appears at relatively small temperature ($k_{B}T\sim \mu
$) in comparison to the homogeneous system \cite{Pitaevskii97PLA}
where this regime occurs at $k_{B}T\gg \mu $. For the highly
elongated condensate of Ref.~\cite{Chevy01}, it has been measured
at $k_{B}T\simeq 0.7 \mu $ a quality factor \cite{quality} for the
transverse quadrupole mode $Q=\Omega _{Q}/(2 \gamma _{Q}^{{\rm
exp}})\sim 200$ which gives a damping rate $\gamma _{Q}^{{\rm
exp}}\sim 10^{-2}\omega _{\perp }$. This value is in agreement
with our calculations of the Landau damping for this mode (see
Fig.~1). A more quantitative comparison is not possible at present
due to the absence of detailed measurements of the damping of this
mode. However, it is reasonable to conclude that the transverse
quadrupole mode in elongated condensates decays via Landau
damping.
It is worth noting that the order of magnitude of the damping rate
of the transverse quadrupole mode is the same as the one
previously estimated for a uniform gas
\cite{Pitaevskii97PLA,Liu1,Liu2,Giorgini1}, for spherical traps
\cite{Guilleumas00} and for anisotropic traps
\cite{Fedichev1,Fedichev2} as well.

The Landau damping for other modes can be calculated analogously.
However, the physical situation for the decay of the singular
transverse monopole mode turns out to be much more complicated.
Formally the calculations can be committed in the same way. The
quantum numbers for this mode are $k=0$, $m=0$ and $n=1$. The functions $u_{%
{\rm osc}}$ and $v_{{\rm osc}}$ depend only on the radial coordinate $%
r_{\perp }$, and the matrix element $A_{ij}$ couples only energy levels ($%
i,j$) with the same quantum numbers $m$ and $k$. That is, the selection
rules corresponding to transverse monopole-like transitions are $\Delta k=0$
(as we have already discussed in the previous section) and $\Delta m=0$.

From the spectrum of elementary excitations, we have found that
there are only $16$ pairs of levels that verify the transverse
monopole selection rules and the energy condition
(\ref{selection}) at $\tilde{k}\neq 0$. The damping strengths
associated to the allowed transitions are smaller than the $\gamma
_{ij}$ values associated to quadrupole-like transitions. As a
result from our calculations the damping occurs quite small. At
$T\simeq 1.5\mu $ we obtain a Landau damping $\gamma_{M} \simeq 7
\times 10^{-5}\omega _{\perp }$ that is one order of magnitude
smaller than the experimental decay \cite{quality} $\gamma
_{M}^{{\rm exp}}\simeq 6 \times 10^{-4}\omega_{\perp }$ measured
in \cite{Chevy01}. Thus, Landau damping cannot explain the
experimentally observed decay of the transverse breathing mode.

Actually, there are reasons to believe that even this small value
of Landau damping obtained in the present calculations, could be
larger than the true one. The point is that in our calculations,
based on perturbation theory, the thermal cloud is assumed to be
in the state of thermodynamic equilibrium and thus one can neglect
its motion. This is correct for the case of quadrupole
oscillations, where the frequencies of the collective motion of
the condensate and the cloud are different. However, due to the
unique peculiar nature of the 2D breathing mode, the
breathing oscillation of the thermal cloud has the same frequency $%
2\omega _{\perp }$ as the condensate. Then, if the oscillation is
excited by the deformation of the trap, as it takes place in the
present experiment \cite{Chevy01}, the cloud will oscillate in
phase with the condensate what will further decrease the Landau
damping \cite {Footnote}. Therefore, this quasi-2D monopole mode
in highly elongated condensates may decay via another damping
mechanism, different from the Landau damping.

After finishing this work it has appeared a preprint by Jackson
and Zaremba \cite{Zaremba02} in which the effect of the collisions
between elementary excitations has been taken also into account.
This effect is usually small in comparison with the Landau
damping. Nevertheless, for this peculiar mode the collisions seems
to be the dominant damping mechanism, yielding a decay rate that
it is in agreement with the experimentally measured value
\cite{Chevy01}. Note, however, that the experiments are produced
at quite high amplitude of oscillations and non-linear effects can
play an important role. For example, as it was noted in
\cite{PitStr}, the transverse breathing mode exhibits a
''parametric'' instability due to the decay into two or more
excitations with non-zero momentum along the axis. In
Ref.~\cite{Kagan01} it has been considered the process in which
the transverse breathing oscillation decays into two excitations
propagating along $z$ with momentum $k$ and $-k$. This damping
mechanism is active also at zero temperature. But in this case the
damping rate depends on the amplitude of the breathing
oscillation.

\section{Summary}

We have investigated the decay of low-lying transverse
oscillations of a large cylindrical condensate. First of all, we
have calculated the transverse normal modes, as well as the
excitations with finite $k$, of the condensate by solving the
linearized time-dependent Gross-Pitaevskii equation. And then,
within the formalism of Ref.~\cite{Guilleumas00}, we have
calculated numerically the matrix elements associated with the
transitions between excited states allowed by the selection rules
of the transverse collective modes. Within a first-order
perturbation theory and assuming the thermal cloud to be in
thermodynamic equilibrium, we have studied the Landau damping of
transverse collective modes due to the coupling with thermal
excited levels as a function of temperature. For the damping rate
of the transverse quadrupole mode, we have found a good agreement
with the experimentally measured value of Ref.~\cite{Chevy01}.
Whereas our result for the transverse breathing mode is one order
of magnitude smaller than the experimental decay. One can conclude
that in a highly elongated cylindrically symmetric condensate,
transverse quadrupole oscillations decay via Landau damping
mechanism, but the transverse breathing mode which has an
anomalously small measured damping rate, may decay via other
damping mechanisms.

\section{Acknowledgments}

We thank J.~Dalibard, G.~Shlyapnikov and E.~Zaremba for useful
discussions. M.~G. thanks the "Ramon y Cajal Program" of the MCyT
(Spain) for financial support.


\begin{references}
\bibitem{quality} In Ref.~\cite{Chevy01} the definition of the
damping rate contains a factor 2 with respect to our definition.
See Eq.~(\ref{dampdef}).
\bibitem{Footnote}  Note that the ''perturbative'' calculations of the
Landau damping can correspond to a different type of experiment,
in which the oscillations of the condensate are excited by the
action of a thin laser beam on the condensate, but not on the
thermal cloud. In this case the ''perturbative'' calculations will
describe the initial stage of the process, when the cloud is still
at rest. As it was presented in the lecture by E. Zaremba
\cite{ZarembaL}, the Landau damping corresponding to this type of
experiment was calculated using the method developed in
Ref.~\cite{Zaremba02} in an elongated condensate with the same
experimental parameters of Ref.~\cite{Chevy01}. However, they have
obtained larger values for the Landau damping of the transverse
breathing mode in comparison with our results. This discrepancy
could be related to the semiclassical nature of their method
\cite{Zaremba02}, which does not take into account the discrete
nature of the energy spectrum, and also to the fact that they
consider a cigar-shaped condensate instead of an infinite cylinder
as in the present calculations. The finite value of the
longitudinal frequency $\omega_{z}$ could be important for this
transverse breathing mode. Therefore, this is still an interesting
open problem.
\end{references}
\end{document}